# Special relativity in the electromagnetic wave

Bernhard Rothenstein and Ioan Damian, Politehnica University, Physics Department. Timisoara, Romania

*Abstract*

*Invariance of the counted number of photons and the Lorentz-Einstein transformations enable us to derive transformation equations for the physical quantities introduced in order to characterize energy emission and transport in a plane and in a spherical electromagnetic wave propagating in vacuum.*

## 1. Introduction

An observer located in an electromagnetic wave and equipped with adequate measuring and detecting devices measures the energy the wave carries, the time interval during which it is emitted, the volume in which the energy is stored, the solid angle inside which the energy is emitted and the surface on which the energy is incident or where from it is emitted. Combinations of these physical quantities lead to other physical quantities introduced in order to characterize emission and reception of radiant energy.[1] Considering measurements performed from different inertial reference frames in relative motion, we should derive transformation equations for each of them.[2,3,4,5] Doing so, we are tempted to use the formulas that account for the time dilation and for the length contraction relativistic effects. The purpose of our paper is to show that these equations lead to an embarrassing paradox: the counted number of stable particles (photons) would not be a relativistic invariant.[6,7,8] Deriving transformation equations we make use of the principle of relativity and of an important consequence of it according to which the counted number of photons is a relativistic invariant.

## 2. Special relativity in the plane electromagnetic wave

In order to keep the problem as simple as possible we first consider the case that involves only measurements of lengths and time intervals. The involved inertial reference frames are K(XOY) and K'(X'O'Y'). Theirs corresponding axes are parallel to each other with the OX(O'X') axes overlapped and K' moves with constant velocity $V$ in the positive direction of the overlapped axes. At the origin of time in the two reference frames ($t=t'=0$) theirs origins are instantly located at the same point in space. The participants in the experiment are the observers $R_i(x_i,0)$ at rest in the K frame and the observers $R'_i(x'_i,0)$, at rest in the K' reference frame, all located in the same monochromatic plane wave propagating with the invariant velocity $c$ in the positive direction of the overlapped axes. Considering this, the electromagnetic oscillations that take in the wave are periodic (harmonic) and the observers of the two reference frames could measure theirs periods T and T' respectively. We could even materialize a clock by erecting a metallic antenna perpendicular to the direction in which the wave propagates. Under the influence of the electric field, the free electrons perform forced oscillations and a potential difference between the ends of the antenna appears. Its period is equal to that of the electromagnetic oscillations taking place in the wave. The observers could also count the number of photons N = N' and could calculate the corresponding frequencies ($f = \frac{1}{T}; f' = \frac{1}{T'}$). The period of the oscillations could serve as a unit of time and observers $R_i$ and $R_i'$ using them as clocks. The period of the oscillations could serve as a unit of time and observers $R_i$ and $R_i'$ use them as clocks. T and T' measured by stationary observers are proper time intervals and are related by the Doppler formula.[9]

$$T = \sqrt{\frac{1-\beta}{1+\beta}} T' = k^{-1} T' \tag{1}$$

where $\beta = \frac{v}{c}$ and $k^{-1} = \sqrt{\frac{1-\beta}{1+\beta}}$. We have

$$f = kf'. \tag{2}$$

Consider that an observer $R_0'(0,0)$ located at the origin O' of his rest frame K' handles a shutter of surface *A'* located in the *Y'O'Z'* plane. He removes it when he receives a wave crest and reinserts it when he receives a second wave crest i.e. for a proper time interval *T'*. The front of the wave created this way advances with

$$\lambda' = cT' \tag{3}$$

and generates a **light volume**

$$V' = AcT' \tag{4}$$

as we show in Figure 1.

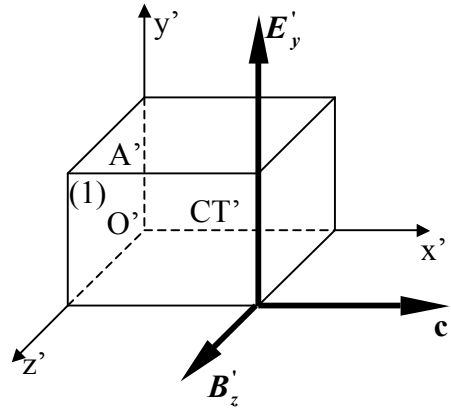

*Figure 1. A shutter of surface A' removed and reinserted in the way of the plane electromagnetic wave generates a "light volume".*

The wavelength $\lambda'$ could serve as an unit of length for the observers at rest in the K' reference frame. During the time interval T', N' = N photons enters the volume *V'* and are uniformly distributed inside it. Because the "light volume" moves with the same velocity as the photons do, the number of photons confined in it does not change with time. The same volume measured by observers of the K frame is

$$V = AcT. \tag{5}$$

Because *A=A'* as surfaces limited by segments perpendicular to the direction of the relative motion we obtain that the volume transforms as

$$V = V' \frac{T}{T'} = k^{-1} V'. \tag{6}$$

$Q = hf$ and $Q' = hf'$ are the energies of the same photon in K and K' respectively. The invariance of the Planck constant yields that

$$Q = kQ'. \tag{7}$$

We can now derive transformation equations for the physical quantities introduced in order to characterize the emission and the reception of energy. The energy stored in the volume V is

$$W = NQ \tag{8}$$

whereas the energy stored in the volume V' is

$$W' = NQ'. \tag{9}$$

The **radiant energy** density defined as energy stored in the unit volume is

$$w' = \frac{W'}{V'} = \frac{NQ'}{V'} \tag{10}$$

in K' and

$$w = \frac{W}{V} = \frac{NQ}{V} \tag{11}$$

in K. Therefore it transforms as

$$w = k^2 w'. \tag{12}$$

The **radiant power** (**radiant energy flux**) defined as energy emitted during unit proper time interval is

$$P = \frac{W}{T} = \frac{NQ}{T} \tag{13}$$

in K and

$$P' = \frac{W'}{T'} = \frac{NQ'}{T'} \tag{14}$$

in K'. It transforms as

$$P = k^2 P'. \tag{15}$$

The pressure exerted by the electromagnetic radiation incident on a black surface is

$$p = \frac{c^{-1}\dfrac{W}{T}}{A} \tag{16}$$

in K and

$$p' = \frac{c^{-1}\dfrac{W'}{T'}}{A'} \tag{17}$$

in K'. Therefore, it transforms as

$$p = k^2 p'. \tag{18}$$

Observers R and R' instantly located at the same point of the OX(O'X') axis detect an electric field $E_y(E'_y)$ respectively and a magnetic induction $B_z(B'_z)$ respectively related to the radiant energy density as[10]

$$w = \varepsilon_0 E_y^2 = \frac{B_z^2}{\mu_0} \tag{19}$$

and as

$$w' = \varepsilon_0 E_y'^2 = \frac{B_z'^2}{\mu_0} \tag{20}$$

from where we obtain the transformation equations

$$E_y = kE'_y \tag{21}$$

$$B_z = kB_z' \tag{22}$$

The electric field and the magnetic induction are perpendicular to the direction of propagation and perpendicular to each other.

We underline the important part played by the Doppler factor k in the transformation equations. Following the same procedure, we can derive transformation equations for all the physical quantities introduced for characterizing the electromagnetic radiation and its actions.

**Special relativity in the spherical wave**

Consider a point like source of radiation S' located at the origin O' of its rest frame K' (Figure 2). It emits uniformly in all directions in space. Consider one of its rays that propagates along a direction, at an angle $\theta'$ with the O'X' axis. The source emits during a very short proper time interval $dT'$ (which is equal to the period of the electromagnetic oscillations taking place in the wave) an energy $dW'$ and $dN'$ photons inside a solid angle $d\Omega'$ centered along the direction $\theta'$. After a time $t'$ of propagation the considered ray generates the event $M'(r'\cos\theta', r'\sin\theta', \frac{r'}{c}) = M'(ct'\cos\theta', ct'\sin\theta', t')$, where $r'$ and $\theta'$ represent the polar coordinates of the point where the event takes place. The Lorentz-Einstein transformations relate the space-time coordinates of events M and M' as

$$x = \gamma r'(\cos\theta' + \beta) \tag{23}$$
$$y = r'\sin\theta' \tag{24}$$
$$r = \gamma r'(1 + \beta\cos\theta') = Dr' \tag{25}$$
$$t = \gamma t'(1 + \beta\cos\theta') = Dt'. \tag{26}$$

Relation
$$\cos\theta = \frac{\cos\theta' + \beta}{1 + \beta\cos\theta'} \tag{27}$$

establishes a relationship between the polar angles describing the aberration of light effect.

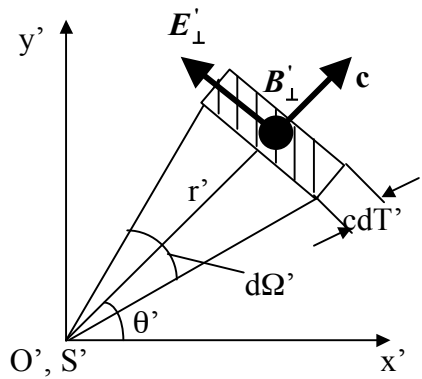

*Figure 2. A point like source of radiation located at the origin O' of its rest frame K' emits energy inside the solid angle $d\Omega'$. The emission starts at t'=0.*

The energy $dW'$ emitted during the proper time interval $dT'$ is confined after a time t' of propagation inside the volume
$$dV = cr^2 d\Omega dT \tag{28}$$
in the K frame and

$$dV' = cr'^2 d\Omega' dT' \tag{29}$$
in the K' frame.

The geometric locus of the events M' is a sphere of radius r' whereas the geometric locus of the events M' is an ellipsoid of rotation. The first has its center in O' whereas the second one has its focal point at O both located at the same point in space at the time $t=t'=0$.[11].

Let $\Omega'$ and $\Omega' + d\Omega'$ be centered along the overlapped axes OX(O'X'). Theirs intersections with the sphere and with the ellipsoid of rotation generate the surfaces

$$dS' = 2\pi r'^2 \sin\theta' d\theta' = r'^2 d\Omega' \tag{30}$$
in K' and
$$dS = 2\pi r^2 \sin\theta d\theta = r^2 d\Omega \tag{31}$$
in K. Relation (30) leads to
$$d\Omega' = 2\pi \sin\theta' d\theta' \tag{32}$$
and from relation (31) we obtain
$$d\Omega = 2\pi \sin\theta d\theta \tag{33}$$

Differentiating both sides of relation (27) we obtain
$$\sin\theta d\theta = D^{-2} \sin\theta' d\theta'. \tag{34}$$

Combining relations (32),(33) and (34) we obtain that the solid angle transforms as
$$d\Omega = D^{-2} d\Omega' \tag{35}$$
and we gain an important geometric invariant in the space where a spherical wave propagates
$$r^2 d\Omega = r'^2 d\Omega'. \tag{36}$$

In accordance with the Doppler Effect formula in the case of this scenario, the energy of a photon transforms as[9]
$$Q = DQ' \tag{37}$$
the period of the electromagnetic oscillations transforms as
$$dT = D^{-1} dT' \tag{38}$$
the volume inside which the radiated energy is confined transforms as
$$dV = \frac{r^2 d\Omega dT}{r'^2 d\Omega' dT'} = D^{-1} dV'. \tag{39}$$

The **radiant energy density** is
$$w = \frac{QdN}{dV} \tag{40}$$
in K and
$$w' = \frac{Q' dN}{dV'} \tag{41}$$
therefore it transforms as
$$w = D^2 w'. \tag{42}$$

The **radiant intensity** defined as energy emitted during the unit proper time interval inside the unit solid angle is
$$I = \frac{\partial^2 N}{dTd\Omega} Q \tag{43}$$
in K and

$$I' = \frac{\partial^2 N}{dT' d\Omega'} Q' \tag{44}$$

consequently it transforms as

$$I = D^4 I'. \tag{45}$$

The electric and the magnetic components of the electromagnetic field are normal to the direction of propagation and normal to each other $E_\perp$ ($B_\perp$) and $E'_\perp$ ($B'_\perp$) in K'. Relations

$$w = \varepsilon_0 E^2 = \frac{B^2}{\mu_0} \tag{46}$$

and

$$w' = \varepsilon_0 E'^2 = \frac{B'^2}{\mu_0} \tag{47}$$

relate them to the radiant energy therefore they transform as

$$E = DE' \tag{48}$$
$$B = DB'. \tag{49}$$

The derivations performed above illustrate the important part played by the Doppler factor in the transformation process.

**Conclusions**

The transformation equations we have derived ensure the relativistic invariance of the counted number of photons a fundamental requirement of the relativistic principle. They hold in in the case of the spherical electromagnetic wave only if the period of the electromagnetic oscillations is very small and the emission of energy starts when the origins of the two frames are momentarily located at the same point in space at the origin of time (*t=t'=0*). An important conclusion is that in order to obtain consistent results in the electromagnetic field we should avoid length contraction that leads to the paradox mentioned in the introduction.[12]

[9]Bernhard Rothenstein, Ioan Damian and Corina Nafornita, "Relativistic Doppler Effect free of "plane wave" and "very high frequency assumptions," Apeiron **12,** 122-135 (2005)

[10]Marcelo Alonso and Edward J.Finn, *Fundamental University Physics II Fields and Waves* (Addisson-Wesley Publishing Company 1975) p.583

[12]Bernhard Rothenstein and Stefan Gall, "A relativistic diagram displaying true values of observable quantities and its applications." Eur.J.Phys. **17**, 71-75 (1196)

[11]W.Geraint V.Rosser, "Comment on "Retardation and relativity: The case of a moving line of charge," by Oleg D. Jefimenko, [Am.J.Phys. **63(5)**, 454-459 (1995)] Am.J.Phys**. 64,** 1202-1203 (1996)
7